# Programmable Floquet Response for Quantum Sensing

**Dionisio Cendoya[1,2], Kang Xu[3], Lisandro Buljubasich[1,2], and Carlos A. Meriles[3,4,†]**

[1]*Facultad de Matemática, Astronomía, Física y Computación, Universidad Nacional de Córdoba, Córdoba, Argentina.*
[2]*CONICET, Instituto de Física Enrique Gaviola (IFEG), Córdoba, Argentina.*
[3]*Department of Physics, CUNY - The City College of New York, New York, NY 10031, USA.*
[4]*CUNY-Graduate Center, New York, NY 10016, USA.*

[†]E-mail: cmeriles@ccny.cuny.edu

Quantum sensing is typically based on phase accumulation and is therefore fundamentally tied to the coherence time of the sensing system. Here we introduce an alternative sensing paradigm based on programmable Floquet locking in periodically driven two-level systems. We show that periodic control generates an effective Floquet locking field whose orientation and amplitude depend on global offset, allowing information to be encoded in the driven population relaxation dynamics, not in the integrated phase. By controlling transfer between distinct Floquet locking manifolds, external perturbations can be mapped onto engineered relaxation responses whose characteristics are programmable through the driving sequence. We numerically demonstrate this approach for both DC and frequency-selective AC sensing, and show that temporal encoding of the response can provide enhanced precision in the presence of slow instrumental fluctuations. These results illustrate the potential of programmable Floquet responses and open the way to exploring new sensing strategies in driven quantum systems.

Quantum sensing with two-level systems is most commonly based on phase accumulation, where an external perturbation is inferred from the coherent evolution of a superposition state[1]. In this paradigm, sensitivity is fundamentally linked to the sensor coherence time, as exemplified by Ramsey interferometry. Dynamical decoupling can substantially extend this time by periodically refocusing environmental fluctuations [2-4], enabling sensitive detection of AC fields matched to the control sequence. This advantage, however, is inherently frequency dependent: slowly varying perturbations remain constrained by the much shorter $T_2^*$, whereas AC sensing can exploit the longer coherence times accessible under dynamical decoupling. This disparity motivates exploring alternative sensing strategies in which information is encoded without phase accumulation, potentially opening access to dynamical timescales beyond those imposed by coherence.

Under resonant driving, a two-level system can be brought into a spin-locked regime in which its dynamics are governed by an effective quantization axis in the rotating (Rabi) frame. The corresponding eigenstates define the basis in which the system's populations are established, with the quantization axis set by the applied control. Because the orientation and amplitude of this quantization axis depend on detuning, spin locking provides a potential mechanism by which external perturbations can be transduced into measurable changes in the system's response[5-11]. Conventional spin locking, however, provides relatively limited control over the dependence of these dynamics on the parameter being sensed. This raises the possibility that more general periodic driving could be used not to protect the sensor, but to engineer how external perturbations are mapped onto its population dynamics.

Recent work has shown that multi-pulse manipulation combined with controlled detuning can generate dynamic spin-locking fields whose magnitude and orientation are determined by the pulse sequence[12,13]. In this setting, the effective quantization axis is no longer fixed by simple continuous driving, but can be shaped in a programmable manner via the chosen control protocol. This capability can be further expanded by the growing use of machine-learning techniques, particularly reinforcement learning, which is being exploited for the automated discovery of pulse sequences that realize targeted dynamical responses, greatly enhancing the accessible control landscape[12,14].

Here, we leverage Floquet quantum control to develop a sensing framework based on engineering the system's response. Rather than encoding information through phase accumulation, the proposed approach exploits relaxation toward a Floquet-locked state, turning decoherence — traditionally regarded as the principal limitation of quantum sensing — into a key component of the sensing mechanism itself[15]. We show that repeated switching between distinct Floquet Hamiltonians converts small transfer mismatches into cumulative population contrast over the $T_1$-like lifetime of the locked state, while periodic modulation of the switching sequence naturally extends the same framework to frequency-selective AC sensing. Lastly, we demonstrate this strategy offers interesting possibilities for parameter estimation, particularly when system fluctuations occur on timescales longer than the driven dynamics.

*Conceptual framework.* We illustrate the basic idea in Fig. 1. The system is subjected to a periodic control sequence consisting of a finite block of pulses that is repeated multiple times. The pulses within each block can, in general, have different phases, amplitudes, frequencies, and non-uniform spacings, providing a high degree of flexibility in the design of the control. A key requirement is the periodic repetition of the block, which establishes a stroboscopic evolution and defines an effective driven frame for the system.

More formally, we write the system Hamiltonian as

$$\mathcal{H}^* = \omega(t) I_z + \mathcal{H}^*_{\mathrm{rf}}(t) + \mathcal{H}^*_{\mathrm{int}}(t)\,, \qquad (1)$$

where $I_z$ is the spin projection operator along $z$, $\mathcal{H}^*_{\mathrm{int}}(t)$ denotes a (secular) interaction term, $\mathcal{H}^*_{\mathrm{rf}}(t)$ encodes the time dependence of the control field amplitude as well as its phase and detuning $\omega(t)$ relative to a reference $\omega_0$ (the two-level system resonance frequency); the asterisk indicates a rotating frame representation. The evolution operator governing the system dynamics writes[16]

$$\mathcal{U}(t) = \mathcal{P}(t)\mathcal{Q}(t)\mathcal{S}(t)\,, \qquad (2)$$

where $\mathcal{P}(t) = \mathcal{T}\exp\left(-i\int_0^t dt' \mathcal{H}^*_{\mathrm{rf}}(t')\right)$ captures the effect of the control, $\mathcal{Q}(t) = \mathcal{T}\exp\left(-i\int_0^t dt'\, \omega(t)\tilde{I}_z(t')\right)$ is an offset-dressed propagator governed by $\tilde{I}_z(t') = \mathcal{P}^\dagger(t) I_z \mathcal{P}(t)$, and $\mathcal{S}(t) = \mathcal{T}\exp\left(-i\int_0^t dt' \tilde{\tilde{\mathcal{H}}}^*_{\mathrm{int}}(t')\right)$ is the evolution under the transformed interaction Hamiltonian $\tilde{\tilde{\mathcal{H}}}^*_{\mathrm{int}}(t') = \mathcal{Q}^\dagger(t)\mathcal{P}^\dagger(t)\mathcal{H}^*_{\mathrm{int}}(t)\mathcal{P}(t)\mathcal{Q}(t)$; lastly, $\mathcal{T}$ is the time-ordering operator. Under periodic driving, the Floquet theorem allows us to write $\mathcal{Q}(t) = \mathcal{F}(t)e^{-i\mathcal{H}_{\mathrm{F}}t}$ with $\mathcal{F}(t)$ denoting a periodic operator chosen so that $\mathcal{F}(kt_c) = 1$ at multiples $k = \{0,1,\ldots\}$ of the cycle time $t_c$, and $\mathcal{H}_{\mathrm{F}} = \mathbf{\Omega}\cdot\mathbf{I}$ is the Floquet Hamiltonian. For a sufficiently strong Floquet field $\mathbf{\Omega}$, Eq. (2) takes the simplified form

$$\mathcal{U}(t_k) = e^{-i\mathcal{H}_{\mathrm{F}}t_k}\mathcal{S}(t_k) \sim e^{-i\mathcal{H}_{\mathrm{F}}t_k}\,, \qquad (3)$$

where we assumed the pulse protocol to be cyclic, i.e., $\mathcal{P}(t_k) = 1$. This effective description defines a quantization axis, analogous to conventional spin locking, but now governed by the structure of the pulse block. Because the control sequence is fixed, variations in external parameters modify the system's resonance condition and therefore the global detuning experienced during the evolution. As a result, the effective quantization axis — and more generally the driven dynamics — acquires a dependence on ω, in turn, a function of a parameter of interest Π. The sensing strategy is to engineer this dependence so that small changes in Π produce large variations in the system's response.

*Quantum sensing via changes in the locking direction.* In general, periodic driving can be engineered so that external perturbations modify either the orientation or the amplitude of the effective Floquet locking field, providing distinct mechanisms for transducing the parameter of interest into a measurable response. While both sensing modalities arise naturally within the present framework, we focus here on the former, as it provides the simplest illustration of the underlying principles. In this regime, the measured signal is determined by the projection of the prepared state onto the corresponding $\omega$-dependent quantization axis. For a fixed initial state and measurement axis $\hat{\boldsymbol{\theta}}$, the dynamic locking signal takes the generic form $S_{\mathrm{DL}}(\omega) \propto \hat{\mathbf{\Omega}}(\omega)\cdot\hat{\boldsymbol{\theta}}$ where $\hat{\mathbf{\Omega}}$ denotes the direction of the locking field. The sensitivity to variations of a parameter Π is therefore determined by the rate at which this direction changes with detuning, i.e.,

$$\eta_\Pi \sim \sigma_{\mathrm{S}}\left(\left|\frac{dS_{\mathrm{DL}}}{d\omega}\right|\left|\frac{d\omega}{d\Pi}\right|\right)^{-1}\sqrt{T_{\mathrm{r}}}\,, \qquad (4)$$

where $\sigma_{\mathrm{S}}$ denotes the experimental noise, and $T_{\mathrm{r}}$ indicates the duration of a single experimental run. Equation (4) highlights the key figure of merit $\left|\frac{dS_{\mathrm{DL}}}{d\omega}\right|$, which is determined by the driven dynamics and can therefore be engineered through the control sequence.

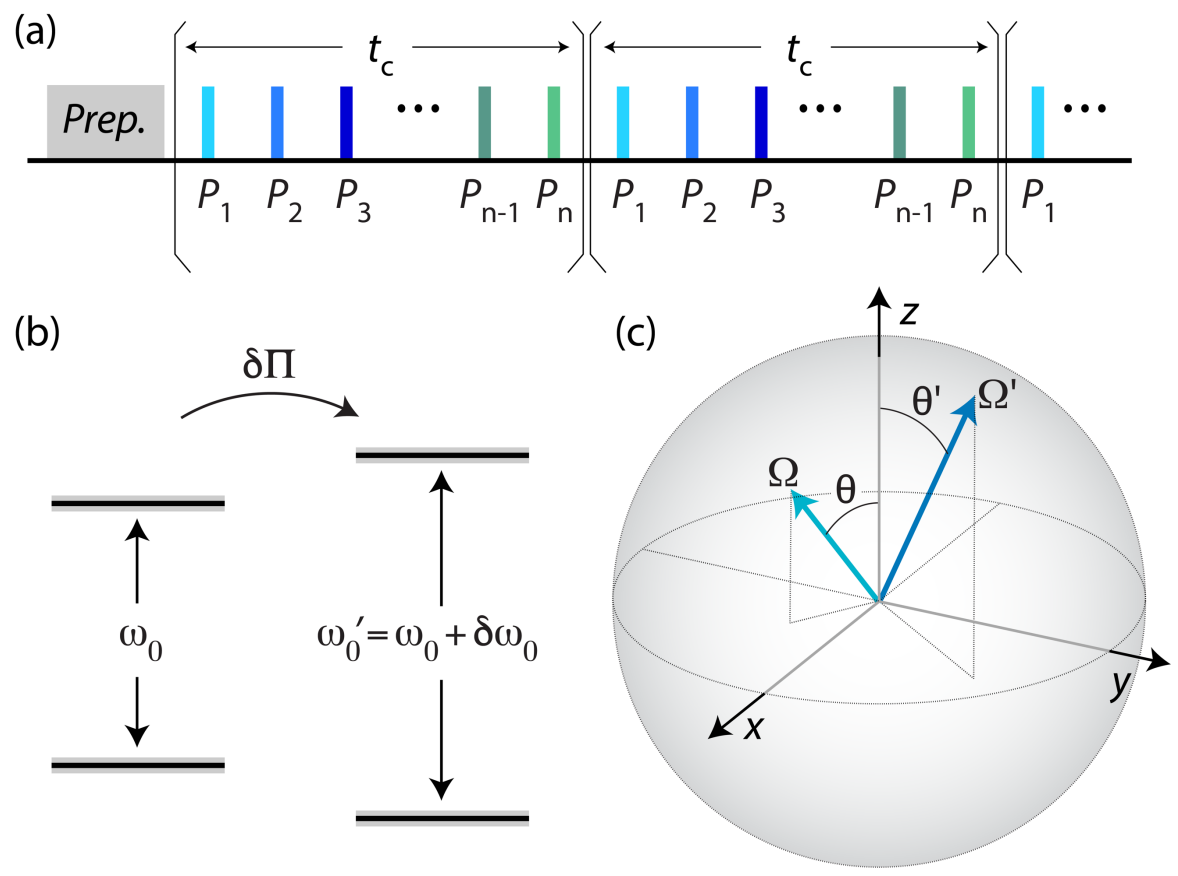

**FIG. 1: Parameter sensing through dynamic spin locking.** (a) A periodically driven two-level system is subjected to a repeated control pulse block $P_1 \ldots P_n$ with cycle time $t_c$, following state preparation. (b) The two-level system has a resonance frequency $\omega_0$, itself a function of an external or internal parameter Π. A change $\delta\Pi$ shifts the resonance to $\omega_0'$, thereby changing the effective detuning $\omega$ experienced under the fixed pulse sequence. (c) Because the periodic drive generates a dynamic spin-locking field Ω(ω) that depends on detuning, the parameter-induced shift in $\omega_0$ modifies both the field direction and amplitude. This, in turn, leads to changes in the amplitude of the locked magnetization, thus altering the observable signal.

*Cumulative sensing via lock-transfer mismatch.* The approach to sensing discussed above relies on the establishment of a detuning-dependent locked state over a (dissipative) settling time, after which the resulting magnetization is projected onto the measurement basis. A convenient extension consists in repeatedly transferring the system between distinct locking conditions in a manner that is exact only for a nominal Hamiltonian. Consider two Floquet Hamiltonians, $\mathcal{H}_F^{(A)}$ and $\mathcal{H}_F^{(B)}$, associated with effective locking axes $\widehat{\mathbf{\Omega}}_A$ and $\widehat{\mathbf{\Omega}}_B$, respectively. The system is first allowed to settle into the locked manifold associated with $\mathcal{H}_F^{(A)}$, after which a control operation $\mathcal{P}_{AB}$ transfers the population to the locking basis of $\mathcal{H}_F^{(B)}$. The transfer is designed such that, in the absence of a parameter variation, $\mathcal{P}_{AB}\left(\widehat{\mathbf{\Omega}}_A(\Pi_0)\right) = \widehat{\mathbf{\Omega}}_B(\Pi_0)$ for a reference parameter value $\Pi_0$.

We can purposely design the transfer so that when the parameter shifts by a small amount $\delta\Pi$, the mapping between one locking basis onto the other is no longer perfect. Under these conditions, the resulting mismatch leaves a component transverse to the new locking axis, which relaxes during the subsequent settling stage. The residual mismatch after the transfer can be characterized by an angle $\epsilon(\delta\omega) \approx \left|\delta\omega\left(\partial_\omega \widehat{\mathbf{\Omega}}_B - \mathcal{P}_{AB}^{(0)} \partial_\omega \widehat{\mathbf{\Omega}}_A\right)\right|$, where $\mathcal{P}_{AB}^{(0)}$ denotes the ideal transfer operation between the two locking manifolds. Following each transfer, relaxation projects the spin onto the new Floquet-locked state, reducing the locked population by a factor $g(\epsilon(\omega)) \leq 1$. Repeating the process $M$ times yields

$$S_{DL}^{(M)}(\delta\omega) \propto [g(\epsilon(\omega))]^M \,. \quad (5)$$

In practice, the protocol is operated in the presence of a global shift $\Delta\omega$, so that small parameter variations produce an approximately linear change of $\epsilon$ and, consequently, an exponentially accumulating change in the measured signal over successive transfer cycles[16]; the number of transfers is ultimately limited by $T_{1\delta}$, the dynamic spin-locking relaxation time.

To illustrate the proposed sensing framework, we consider a physical system comprising an electronic spin-1/2 **I** acting as a local probe[17,18]. An example of a periodic driving protocol — adapted from the WaHuHa family[19,20] — is shown in Fig. 2a. The sequence consists of two distinct pulse blocks, A and B, each repeated sufficiently long for the spin to relax toward the corresponding Floquet-locked state. For the particular sequences considered here, the locking fields are engineered to lie entirely within the $yz$ plane (insert in Fig. 2a), so that changes $\delta\omega$ in detuning — or, equivalently, $\Pi$-induced changes $-\delta\omega_0$ in the spin resonance frequency — rotate their orientation within this plane in opposite directions. Consequently, a transfer operation that is exact at the nominal detuning becomes progressively imperfect in the presence of an $\omega_0$ shift, leading to an increasing mismatch between the prepared state and the new Floquet-locked state. Figure 2b shows the resulting dependence of the relative angle between the two locking fields on resonance shifts, which forms the basis of the sensing protocol.

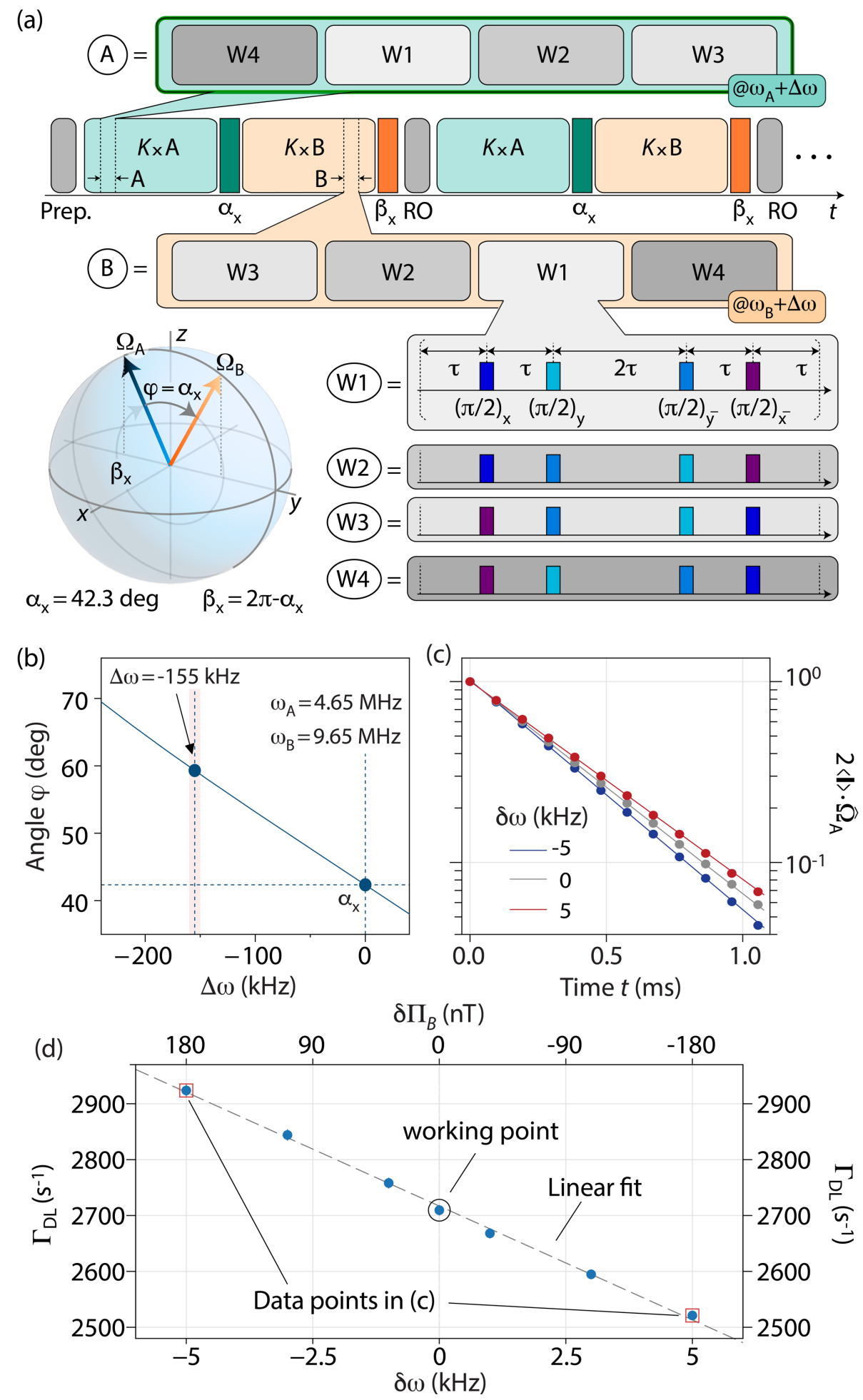

**FIG. 2: Floquet response engineered through periodic driving.** (a) Example periodic pulse sequence. $K$ represents the number of cycles, while $\omega_A + \Delta\omega$, $\omega_B + \Delta\omega$ denote the frequency detuning of all pulses in each block. (Insert) Effective Floquet locking fields associated with successive pulse blocks. (b) Relative orientation of the Floquet locking fields as a function of detuning; here, $\mathcal{P}_{AB}^{(0)} = e^{-i\alpha_x I_x}$. (c) Evolution of the locked-state population for representative resonance shifts. (d) Floquet sensing observable as a function of $\delta\omega$. The upper axis shows an equivalent magnetic-field shift $\delta\Pi_B$ for an electron spin. Prep: Preparation. RO: Readout.

The resulting mismatch is converted into a measurable population contrast through relaxation toward the Floquet-locked states, a process we mimic here by imposing a Lorentzian noise spectral density[16]. Figure 2c illustrates the resulting relaxation dynamics for representative values of the detuning. Following a switch from block A to block B, the population initially prepared in the locked state of A is no longer aligned with the locking field of the new Floquet Hamiltonian. The transverse component generated by this mismatch relaxes, leaving a population — different than the original — projected along the locking axis of block B. Given the built-in dependence on detuning, the resulting population dynamics provide a direct mapping of resonance shifts onto relaxation.

In its simplest implementation, the sensing response can be obtained from a single population measurement after a selected interrogation time. More generally, however, the relaxation trajectory itself can serve as the sensing observable, yielding the time-dependent response $S_{\mathrm{DL}}(t) \propto \exp(-\Gamma_{\mathrm{DL}}(\omega)t)$, from which the corresponding relaxation rate $\Gamma_{\mathrm{DL}}(\omega)$ can be inferred. Note that although the formalism above describes a single two-level system, the same dynamics apply to an ensemble of noninteracting, identically driven spins, where weak population measurements could be realized with sufficiently small backaction on the collective spin polarization. Figure 2d shows the relaxation rates extracted from the population trajectories in Fig. 2c as a function of resonance shift. As discussed later, accessing this full temporal response opens alternative strategies for parameter estimation in the presence of experimental fluctuations.

*Extension to AC sensing.* Consider a parameter modulation of the form $\delta\Pi(t) = \Pi_{\mathrm{ac}}\cos(\omega_{\mathrm{ac}}t + \phi)$, which produces a corresponding time dependence of the effective Floquet locking conditions. As before, each transfer-and-settle step samples the instantaneous value of the perturbation through the mismatch angle, converting it into a small population change. For AC sensing, however, the pulse sequence is organized into supercycles comprising two consecutive A→B transfer cycles with equal and opposite bias detunings about a chosen working point, as illustrated in Fig. 3a. The two supercycles are related by mirror symmetry in a way such that their response to resonance shifts is opposite[16] (Figs. 3b and 3c).

The interval between successive supercycles defines the modulation period with which the protocol samples the external perturbation. Synchronizing this period with the signal frequency produces the frequency-selective response shown in Fig. 3d while preserving the cumulative relaxation mechanism underlying the DC protocol. As in filter-function-based techniques such as CPMG or XY8 magnetometry, periodic modulation gives rise to frequency selectivity. Here, however, the detection frequency is determined by the duration of the Floquet megacycle, making the protocol well suited to comparatively low frequencies (typically below 50−100 kHz), where conventional dynamical-decoupling protocols generally become less effective[16]. Note that because the frequency selectivity is set by the long $T_1$-limited accumulation of population contrast, the resulting response is substantially narrower than the corresponding XY8 filter.

*Sensitivity and long-term precision.* The performance of the relaxation-based protocol must be considered separately in the short- and long-time limits. Under ideal measurement conditions, one can show that extracting a resonance shift from a relaxation trajectory generally carries a statistical penalty relative to conventional Ramsey sensing[16]. This comparison changes, however, in the presence of slow experimental instabilities. As illustrated in Fig. 4a, intensity-based sensing is directly susceptible to fluctuations in signal amplitude and contrast, which can mimic changes induced by the parameter of interest. By contrast, such fluctuations can substantially modify the overall relaxation signal while leaving its characteristic rate unchanged. Provided they remain approximately constant over the duration of each relaxation trace, the rate can therefore be extracted from the temporal shape of the response with reduced sensitivity to run-to-run variations in measurement conditions.

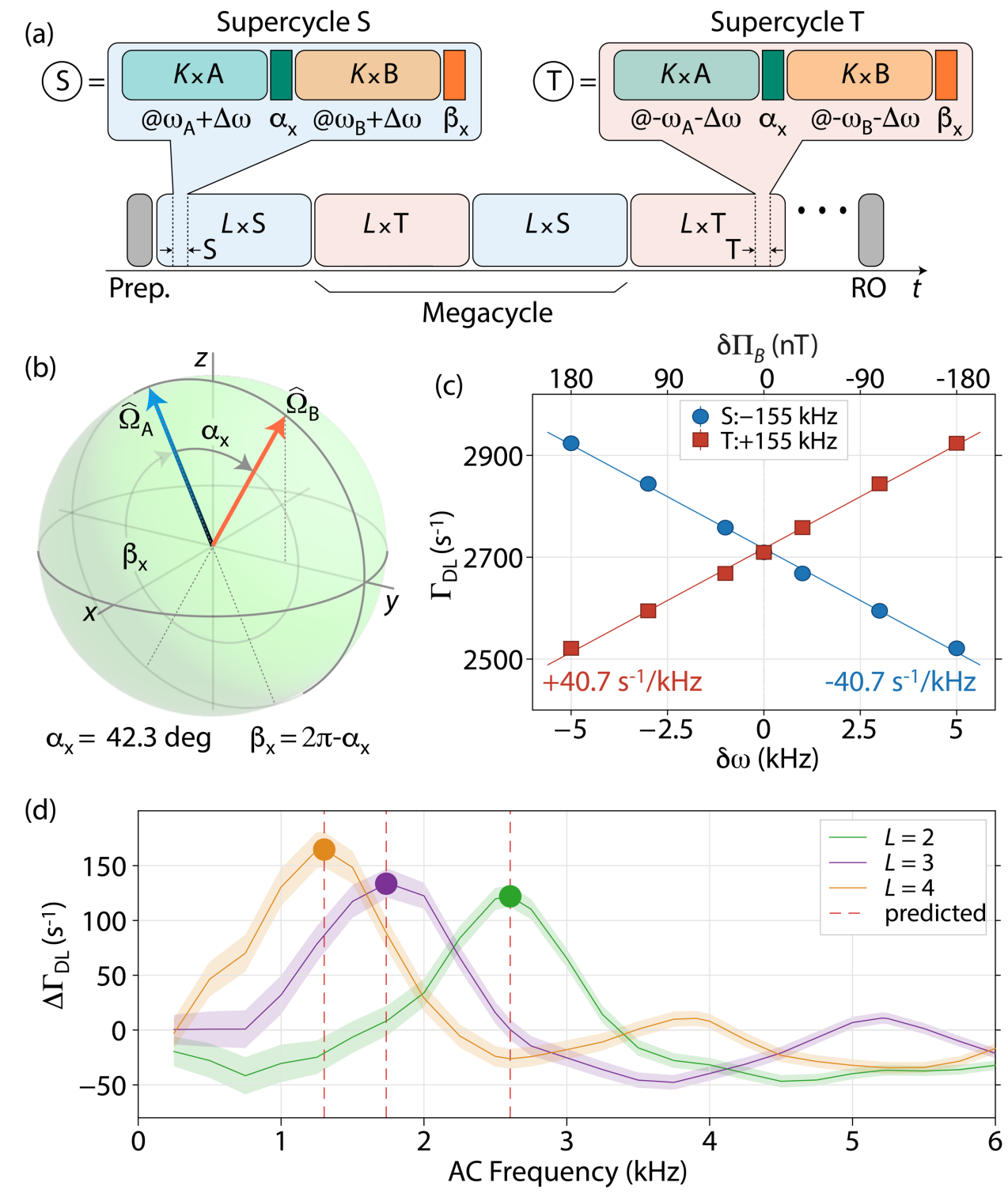

**FIG. 3: AC sensing via Floquet megacycles.** (a) Pulse sequence schematics; $L$ denotes an integer and $\Delta\omega$ is the bias detuning, chosen to maximize the slope. (b) Floquet locking-field geometry. (c) Detuning dependence of the transfer mismatch. (d) Frequency response of the protocol, illustrating selective detection of an oscillatory perturbation.

Figure 4b illustrates this tradeoff for representative system noise containing white and $1/f$ components[16]. At short averaging times, Ramsey provides the smaller minimum resolvable resonance shift, reflecting its superior intrinsic sensitivity. With increasing measurement time, however, correlated system fluctuations — here exemplified through readout contrast fluctuations — cause the Ramsey precision to depart from the usual inverse-square-root scaling, eventually reaching a plateau and deteriorating at longer times. Despite its poorer initial sensitivity, the relaxation-based estimate remains comparatively insensitive to these slow fluctuations and continues to improve, eventually yielding a smaller resolvable resonance shift. This behavior is analogous to that experimentally observed in excited-state-lifetime-based magnetometry[21], and serves to illustrate the broader opportunities enabled by encoding sensing information in programmable relaxation dynamics.

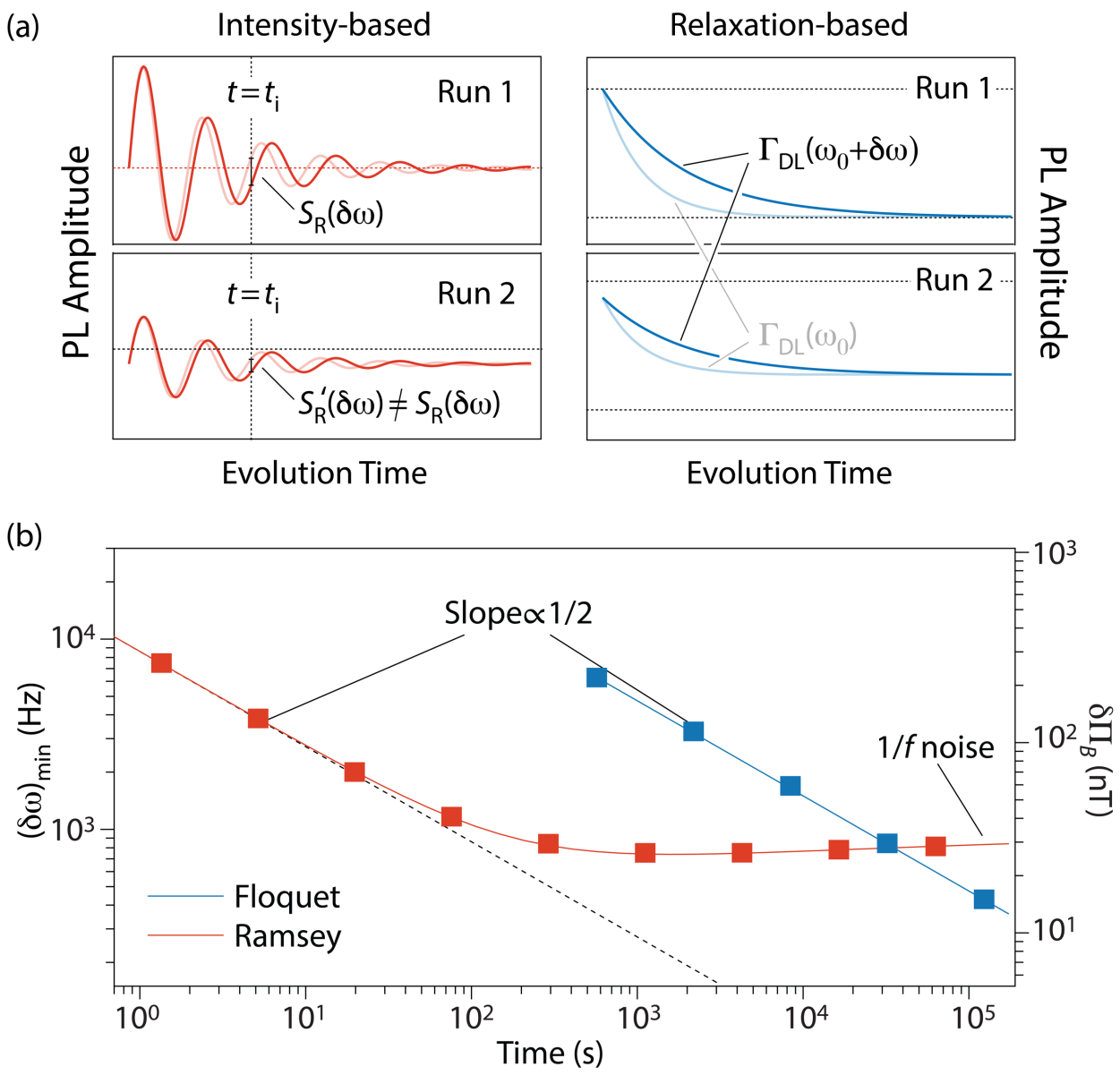

**FIG. 4: Sensitivity and long-term precision.** (a) Comparison of measurement strategies. Unlike an intensity measurement at time $t_i$ (left), the relaxation rate Γ extracted from the temporal response (right) is relatively insensitive to slow changes in overall amplitude, offset, and contrast. (b) Simulated minimum resolvable frequency shift versus measurement time. Despite lower initial sensitivity, Floquet relaxometry maintains improving precision as Ramsey sensing becomes limited by instrumental fluctuations. See Ref. [16] for details.

In summary, we introduced a framework for quantum sensing based on the Floquet response of periodically driven two-level systems. Instead of relying on phase accumulation, the proposed approach encodes information in the relaxation dynamics of Floquet-locked states, providing an alternative route for mapping external perturbations onto measurable population changes. We demonstrated this principle for DC sensing and extended it to frequency-selective AC detection through appropriately designed Floquet transfer sequences. We further showed that extracting information from the temporal relaxation response can trade short-term sensitivity for improved long-term precision in the presence of slow experimental instabilities.

Besides leveraging the Floquet field direction, future work could explore perturbation-induced changes in the Floquet-field amplitude, particularly attractive in interacting ensemble systems where spontaneously generated entanglement could be exploited to convert resonance shifts into altered many-body dynamics. More generally, the ability to engineer Floquet response functions opens a vast control landscape, difficult to explore exhaustively through intuition or analytical design alone[22-25]. The present framework is therefore naturally suited to AI-assisted optimization, where machine-learning algorithms can search the space of periodic driving protocols for sensing dynamics tailored to specific signals, noise environments, and measurement constraints[14]. These results suggest therefore a shift in perspective from designing pulse sequences that preserve coherence to designing driven quantum systems whose response to external perturbations is itself programmable.

We thank Diego Bussandri for useful discussions and assistance with the manuscript preparation. D.C. and L.B. acknowledge support from Secyt-UNC 32520230100029CB and CONICET PIP 11220200102451CO. C.A.M. acknowledges support by the U.S. Department of Energy, Office of Science, National Quantum Information Science Research Centers, Co-design Center for Quantum Advantage (C2QA) under contract number DE-SC0012704; K.X. acknowledges support from the U.S. National Science Foundation via grant NSF-2328993; they also acknowledge access to the facilities and research infrastructure of the NSF CREST IDEALS, grant number NSF-2112550.

# Supplementary Material for

# "Programmable Floquet Response for Quantum Sensing"

**Dionisio Cendoya[1,2], Kang Xu[3], Lisandro Buljubasich[1,2], and Carlos A. Meriles[3,4,†]**

[1]*Facultad de Matemática, Astronomía, Física y Computación, Universidad Nacional de Córdoba, Córdoba, Argentina.*
[2]*CONICET, Instituto de Física Enrique Gaviola (IFEG), Córdoba, Argentina.*
[3]*Department of Physics, CUNY - The City College of New York, New York, NY 10031, USA.*
[4]*CUNY-Graduate Center, New York, NY 10016, USA.*

[†]E-mail: cmeriles@ccny.cuny.edu

## I. Dynamics of periodically driven two-level systems

Periodic driving modifies not only the coherent evolution of a two-level system, but also the manner in which it exchanges energy with its environment. Rather than relaxing toward the eigenstates of the static Hamiltonian, the driven system relaxes toward the eigenstates of an effective Floquet Hamiltonian whose properties are entirely determined by the applied control. Consequently, both the relaxation rates and the steady-state populations become programmable through the driving protocol. In this section we develop a general description of these dynamics, independent of any particular pulse sequence or sensing protocol. The resulting formalism provides the foundation for the relaxation-assisted sensing schemes introduced in the main text.

### I.1 Floquet description of periodically driven two-level systems

We consider a periodically driven two-level system whose dynamics, in the rotating frame, are governed by

$$\mathcal{H}^* = \omega(t) I_z + \mathcal{H}_{\mathrm{rf}}^*(t) + \mathcal{H}_{\mathrm{int}}^*(t)\,, \tag{1}$$

where $I_z$ is the *z*-projection of a spin-1/2 operator, $\mathcal{H}_{\mathrm{rf}}^*(t)$ is a periodic driving Hamiltonian satisfying $\mathcal{H}_{\mathrm{rf}}^*(t+t_{\mathrm{c}}) = \mathcal{H}_{\mathrm{rf}}^*(t)$, and $\mathcal{H}_{\mathrm{int}}^*(t) = \lambda(t) I_z$ where $\lambda(t)$ denotes a stochastic process satisfying $\langle\lambda(t)\rangle = 0$, here characterized by a Lorentzian spectral density $\Lambda(\omega) = \int_{-\infty}^{\infty} d\tau\, e^{i\omega\tau}\langle\lambda(t+\tau)\lambda(t)\rangle = \Lambda_0\, \Upsilon^2/(\omega^2+\Upsilon^2)$. Throughout this work, the detuning $\omega(t)$ — in general, a function of time — represents the instantaneous frequency offset between the driven two-level system and the applied control field. In the present implementation, all pulses within a given Floquet block are applied with a common carrier frequency, so that $\omega(t)$ remains constant throughout each block and changes only between successive blocks. The formalism presented below, however, is not restricted to this case and remains applicable to arbitrary time-dependent detuning profiles.

To separate the dynamics generated by the periodic control from those arising from the resonance offset, we introduce the control propagator

$$\mathcal{P}(t) = \mathcal{T}\exp\left(-i\int_0^t dt'\mathcal{H}_{\mathrm{rf}}^*(t')\right), \tag{2}$$

followed by the offset-dressed propagator

$$\mathcal{Q}(t) = \mathcal{T}\exp\left(-i\int_0^t dt'\,\omega\tilde{I}_z(t')\right), \tag{3}$$

where $\tilde{I}_z(t) = \mathcal{P}^\dagger(t)\tilde{I}_z\mathcal{P}(t)$. This separation is particularly convenient because it isolates the action of the periodic drive from the slowly varying resonance offset, allowing the same formalism to describe both static and time-dependent perturbations. Unlike conventional average Hamiltonian theory, which is valid for sufficiently small resonance offset, the present construction remains applicable as the detuning takes arbitrarily large values.

For a periodic driving sequence, Floquet theory allows the offset-dressed propagator to be decomposed as

$$\mathcal{Q}(t) = \mathcal{F}(t)\exp(-i\mathcal{H}_{\mathrm{F}}t)\,, \tag{4}$$

where $\mathcal{F}(t+t_c) = \mathcal{F}(t)$ is periodic and $\mathcal{H}_{\mathrm{F}}$ is the effective Floquet Hamiltonian. The latter defines an effective locking field whose orientation and magnitude depend parametrically on the instantaneous resonance offset,

$$\mathcal{H}_{\mathrm{F}}(\omega) = \mathbf{\Omega}(\omega)\cdot\mathbf{I}\,. \tag{5}$$

The Floquet Hamiltonian defines an effective quantization axis, $\hat{\mathbf{\Omega}}$, which depends on the applied driving protocol and, in general, on the instantaneous detuning.

The sensing protocols discussed in the main text rely on relaxation toward the instantaneous Floquet-locked state. Throughout this work, we describe this relaxation by explicitly introducing stochastic magnetic-field fluctuations into the system Hamiltonian. This approach allows the dynamics to be simulated for arbitrary environmental spectra while remaining independent of the microscopic origin of the noise. For a given realization of the stochastic field, the spin dynamics are obtained by numerically integrating the time-dependent Schrödinger equation; observable quantities are then computed by averaging over a sufficiently large ensemble of independent noise realizations, from which the Floquet-state populations and spin expectation values are evaluated. Notice that although a Lindblad description is computationally more efficient, we found that accurately reproducing the stochastic dynamics generally requires recalibration of the relaxation parameters as the driving conditions change. We therefore consistently employ explicit stochastic sampling throughout this work, ensuring a uniform treatment of noise across the different sensing protocols.

**I.2 Sensitivity of relaxation-assisted Floquet sensing**

The response of the sensing protocol results from the combined action of two distinct relaxation processes. The first originates from the finite lifetime of the Floquet-locked state and is characterized by the relaxation time $T_{1\delta}$. The second is associated with the imperfect transfer between successive Floquet locking manifolds and depends on the mismatch angle $\epsilon$ between the transferred spin polarization and the locking axis of the subsequent Floquet Hamiltonian. The measured signal therefore reflects the competition between preserving the locked-state population for as long as possible and maximizing the sensitivity of the transfer to changes in the parameter of interest.

The transfer mismatch can be quantified by the angle

$$\epsilon(\omega) = \cos^{-1}\left(\widehat{\boldsymbol{\Omega}}_{\mathrm{B}}(\omega) \cdot \mathcal{P}_{\mathrm{AB}}^{(0)} \widehat{\boldsymbol{\Omega}}_{\mathrm{A}}(\omega)\right) \tag{6}$$

where $\widehat{\boldsymbol{\Omega}}_{\mathrm{A,B}}$ denote the Floquet locking fields associated with the two pulse blocks, and $\mathcal{P}_{\mathrm{AB}}^{(0)}$ represents the ideal transfer operation at a selected point. The pulse sequence is first calibrated at a reference point $\omega_{\mathrm{r}}$, chosen so that the transfer between the two Floquet locking manifolds is exact, i.e., $\epsilon(\omega_{\mathrm{r}}) = 0$. As a result, the transfer efficiency is locally symmetric about $\omega_{\mathrm{r}}$, and the corresponding population response depends only quadratically on small deviations of the resonance offset. To obtain a linear response, the sensing protocol is instead operated about a nearby working point $\omega_{\mathrm{w}}$, where the nominal mismatch assumes a finite value, $\epsilon(\omega_{\mathrm{w}}) = \epsilon_{\mathrm{w}}$. Small parameter-induced variations of the resonance frequency $\delta\omega$ therefore produce

$$\epsilon(\omega) = \epsilon_{\mathrm{w}} - \left.\frac{\partial \epsilon}{\partial \omega}\right|_{\omega_{\mathrm{w}}} \delta\omega\,, \tag{7}$$

leading to a first-order change of the transfer efficiency.

In general, one can determine the optimal working point by first noticing that, for a given population transfer efficiency $g(\epsilon(\omega)) \leq 1$, the cumulative response at a time $t$ is given by

$$S_{\mathrm{DL}}(\omega, t) = [g(\epsilon(\omega))]^{M} \exp\left(-\frac{t}{T_{1\delta}(\omega)}\right) = \exp\left(-2Mf\big(\epsilon(\omega)\big)\right) \exp\left(-\frac{t}{T_{1\delta}(\omega)}\right), \tag{8}$$

where $M$ denotes the number of transfers and we defined $f\big(\epsilon(\omega)\big) = -0.5 \ln g(\epsilon(\omega))$. In the protocols considered here, each A or B block contains $K$ repetitions of the elementary Floquet cycle, with $K$ chosen such that $Kt_c \approx 2T_{2\delta}$. Therefore, the number of transfers accumulated during an interrogation time $t$ is $M \approx t/2T_{2\delta}$. The Floquet-locked population then takes the form

$$S_{\mathrm{DL}}(\omega, t) = \exp\left(-\frac{t}{T_{2\delta}(\omega)} f\big(\epsilon(\omega)\big)\right) \exp\left(-\frac{t}{T_{1\delta}(\omega)}\right). \tag{9}$$

For a measurement limited by an approximately parameter-independent readout noise, the sensitivity per unit bandwidth scales as

$$\eta(\omega, t) \propto \frac{\sqrt{t}}{|\partial_{\omega} S_{\mathrm{DL}}(\omega, t)|} = \frac{e^{\Gamma_{\mathrm{DL}}(\omega) t}}{\sqrt{t}\,|\partial_{\omega} \Gamma_{\mathrm{DL}}(\omega)|}\,, \tag{10}$$

where we defined $\Gamma_{\mathrm{DL}}(\omega) = f\big(\epsilon(\omega)\big)/T_{2\delta}(\omega) + 1/T_{1\delta}(\omega)$. For a given $\omega$, minimization with respect to the interrogation time gives $t_{\mathrm{opt}} = 1/\big(2\Gamma_{\mathrm{DL}}(\omega)\big)$. The corresponding optimized sensitivity therefore scales as

$$\eta_{\mathrm{opt}}(\omega) \propto \frac{\sqrt{\Gamma_{\mathrm{DL}}(\omega)}}{|\partial_{\omega} \Gamma_{\mathrm{DL}}(\omega)|}\,, \tag{11}$$

which, in turn, can be used to determine the working point $\omega_{\mathrm{w}}$. Physically, the optimum reflects a compromise between operating close enough to the reference point to retain a long $T_{1\delta}$, while moving sufficiently far away so that changes in detuning produce a large variation of the mismatch-dependent decay rate. This criterion is used below to determine both the working offset and the corresponding interrogation time for the specific DC and AC protocols.

An important aspect of the optimization is that both relaxation times, $T_{1\delta}$ and $T_{2\delta}$, are themselves functions of the detuning. This dependence originates from the evolution of the Floquet Hamiltonian with resonance offset: Changes in detuning modify both the magnitude and the orientation of the effective Floquet locking field, thereby altering the spectral overlap between the driven spin dynamics and the environmental fluctuations. For the case considered here of Lorentzian noise acting along the laboratory $z$ direction, one generally expects the dynamic spin-locking lifetime $T_{1\delta}$ to reach a maximum at the detuning where the effective locking field is most closely aligned with the noise axis, while the corresponding transverse relaxation time $T_{2\delta}$ reaches a minimum. Numerical calculations confirming this behavior for the pulse sequences employed in this work are presented in the following sections.

### I.3 Optimization for relaxation-rate estimation and long-term precision

The sensitivity analysis above assumes that the sensing information is extracted from the population measured at a selected interrogation time. An alternative implementation is to use the relaxation dynamics themselves as the sensing observable. In this case, the resonance offset is inferred from the relaxation rate $\Gamma_{\mathrm{DL}}(\omega)$, and the relevant optimization problem is no longer determined solely by the slope of the population response. Instead, the protocol must provide sufficient information within each relaxation trajectory to estimate $\Gamma_{\mathrm{DL}}$ accurately, while completing the measurement on a timescale short compared with the instrumental fluctuations that the rate-based measurement is intended to reject.

To formalize this criterion, consider a sequence of weak population measurements acquired at times $t_k$ during a single experimental realization. For an exponential relaxation, the measured signal can be written as

$$S_{\mathrm{DL}} = \mathcal{V} e^{-\Gamma_{\mathrm{DL}}(\omega) t_k} + \mathcal{W} + \varepsilon_k \,, \tag{12}$$

where $\mathcal{V}$ and $\mathcal{W}$ describe the signal amplitude and background, respectively, and $\varepsilon_k$ denotes measurement noise. Importantly, $\mathcal{V}$ and $\mathcal{W}$ need not be stable between successive realizations. Provided that their variations are slow compared with the duration $T_{\mathrm{tr}}$ of an individual relaxation trace, they can be treated as nuisance parameters that are approximately constant within that trace. The resonance offset is then obtained from the fitted rate according to

$$\delta\omega = \frac{\delta\Gamma_{\mathrm{DL}}}{|\partial_\omega \Gamma_{\mathrm{DL}}|_{\omega_{\mathrm{w}}}} \,, \tag{13}$$

so that optimization requires simultaneously maximizing the transduction $|\partial_\omega \Gamma_{\mathrm{DL}}|_{\omega_{\mathrm{w}}}$ and minimizing the statistical uncertainty with which $\Gamma_{\mathrm{DL}}$ can be determined.

The latter condition can be quantified using the Fisher information. For independent Gaussian measurements of variance $\sigma^2$, the Fisher information matrix for the parameters $\boldsymbol{\chi} = (\mathcal{V}, \mathcal{W}, \Gamma)$ is

$$\mathcal{I}_{a,b} = \frac{1}{\sigma^2} \sum_k \frac{\partial S_{\mathrm{DL}}(t_k)}{\partial \chi_a} \frac{\partial S_{\mathrm{DL}}(t_k)}{\partial \chi_b} \,. \tag{14}$$

The minimum variance of an unbiased estimator of the relaxation rate is bounded by

$$\mathrm{Var}(\Gamma_{\mathrm{DL}}) \geq \mathcal{I}_{\Gamma\Gamma}^{-1} \,, \tag{15}$$

where inversion of the full matrix accounts for the fact that $\mathcal{V}$ and $\mathcal{W}$ are not known independently. The corresponding frequency uncertainty is therefore bounded by

$$\mathrm{Var}(\omega) \geq \frac{\mathcal{I}_{\Gamma\Gamma}^{-1}}{|\partial_\omega \Gamma|^2} \,. \tag{16}$$

This expression provides the optimization criterion for rate-based Floquet sensing.

A useful physical interpretation follows directly from the structure of the exponential response. Measurements performed over a time interval much shorter than the relaxation time contain little information about $\Gamma_{\mathrm{DL}}$, since for $\Gamma_{\mathrm{DL}} T_{\mathrm{tr}} \ll 1$, $e^{-\Gamma_{\mathrm{DL}} t} \approx 1 - \Gamma_{\mathrm{DL}} t$ and changes in $\Gamma_{\mathrm{DL}}$ become difficult to distinguish from changes in signal amplitude and offset. Conversely, extending the trace far beyond the relaxation time provides progressively little additional information once the population has approached its asymptotic value. Accurate rate estimation therefore requires the relaxation trajectory to span an appreciable fraction of the decay, with measurements distributed over the corresponding temporal interval. This requirement replaces the simpler criterion of maximizing the population slope at a single interrogation time.

In the present implementation there is an additional constraint on the temporal sampling. Following each weak measurement, the transverse component generated by the readout must be allowed to dissipate before the next population measurement. The spacing $\Delta t$ between successive samples must therefore satisfy, parametrically, $\Delta t \gtrsim T_{2\delta}$ up to a numerical factor determined by the tolerated residual transverse polarization. Consequently, a trace containing $N$ useful samples requires approximately

$$T_{\mathrm{tr}} \gtrsim (N-1)\Delta t \gtrsim (N-1)T_{2\delta}\,. \tag{17}$$

Together with the requirement $\Gamma_{\mathrm{DL}}T_{\mathrm{tr}}\sim\mathcal{O}(1)$, this defines a practical window in which sufficiently many statistically independent points can be acquired while still resolving an appreciable portion of the relaxation.

The second requirement arises from instrumental stability. Rate estimation is advantageous over intensity-based sensing only when fluctuations in amplitude, background, or contrast remain sufficiently slow during an individual trajectory. Denoting by $\tau_{\mathrm{inst}}$ the characteristic timescale over which these quantities change appreciably, the desired regime is

$$T_{\mathrm{tr}} \ll \tau_{\mathrm{inst}}\,. \tag{18}$$

Slow variations may then produce substantial changes between successive traces while leaving the temporal shape, and hence the inferred rate, comparatively unaffected. In contrast, fluctuations with appreciable spectral weight on timescales comparable to or shorter than $T_{\mathrm{tr}}$ distort the relaxation trajectory itself and directly contribute to the uncertainty in $\Gamma_{\mathrm{DL}}$.

The operating conditions for rate-based Floquet sensing are therefore determined by the combined window

$$(N-1)T_{2\delta} \lesssim T_{\mathrm{tr}}\sim\Gamma^{-1} \ll \tau_{\mathrm{inst}}\,, \tag{19}$$

together with a sufficiently large frequency transduction $|\partial_\omega\Gamma_{\mathrm{DL}}|$. The optimum therefore corresponds to a relaxation rate that is sufficiently sensitive to resonance offset, sufficiently slow to permit temporal sampling, but sufficiently fast that the complete sensing information can be acquired within the quasistatic window of the experimental noise.

To gain physical insight into the optimization criterion, it is useful to consider the regime where the mismatch-induced decay dominates the detuning dependence of the response, so that $\Gamma_{\mathrm{DL}}(\omega) \approx f\big(\epsilon(\omega)\big)/T_{2\delta}(\omega)$. Equation (11) then becomes

$$\eta_{\mathrm{opt}}(\omega) \propto \frac{\sqrt{f(\epsilon)T_{2\delta}}}{\left|\partial_\omega f(\epsilon) - \dfrac{f(\epsilon)}{T_{2\delta}}\partial_\omega T_{2\delta}\right|}\,. \tag{20}$$

A further simplification is obtained when the working point is chosen to coincide with the detuning at which $T_{2\delta}$ (and, for the Lorentzian spectra considered here, also $T_{1\delta}$) reaches an extremum, so that $\partial_\omega T_{2\delta} \approx 0$. Under these conditions,

$$\eta_{\mathrm{opt}}(\omega) \propto \frac{\sqrt{f(\epsilon)T_{2\delta}}}{|\partial_\omega f(\epsilon)|}\,. \tag{21}$$

Since the transfer efficiency is roughly approximated by $g(\epsilon) = \cos\epsilon$, one has $f(\epsilon) \approx -0.5\ln(\cos\epsilon) \approx -0.5\ln(\cos\epsilon_{\mathrm{w}}) - 0.5\tan(\epsilon_{\mathrm{w}})\frac{\partial\epsilon}{\partial\omega}\Big|_{\omega_{\mathrm{w}}}\delta\omega$. Substituting this expression into Eq. (21) yields,

$$\eta_{\mathrm{opt}} \propto \frac{\sqrt{f(\epsilon_{\mathrm{w}})T_{2\delta}}}{\left|0.5\tan(\epsilon_{\mathrm{w}})\,\partial_\omega\epsilon|_{\omega_{\mathrm{w}}}\right|}\,. \tag{22}$$

Equation (22) therefore shows that, once the working point is chosen at the relaxation optimum, the sensitivity separates into two largely independent control parameters: the Floquet dynamics, through $T_{2\delta}$, and the transfer response, through $|\partial_\omega\epsilon|$. The former favors operating where $T_{2\delta}$ is shortest — while $T_{1\delta}$ remains maximal — whereas the latter can be enhanced by designing the transfer operation itself to be especially sensitive to resonance offset. In particular, composite transfer pulses may be engineered to remain exact at the calibration point while developing a large $\partial_\omega\epsilon$ at the working point, thereby further amplifying the population response without altering the underlying Floquet locking dynamics.

## II. Design and implementation of the sensing protocol

We now describe the construction of the sensing protocols introduced in the main text, focusing on the physical considerations guiding the selection of the Floquet blocks, transfer operations, and operating conditions. Rather than pursuing an exhaustive optimization over the space of possible control sequences, we develop a representative implementation. This choice allows us to exploit known properties of dynamic spin locking while illustrating the design principles that could subsequently guide a more systematic search using reinforcement learning.

### II.1 Selection and characterization of the Floquet locking blocks

The starting point is the identification of two periodically driven configurations suitable for repeated population transfer. In general, each periodic sequence generates a characteristic detuning-dependent Floquet field, together with longitudinal and transverse relaxation times determined by the coupling to the environment. A useful sensing protocol requires not only that these quantities exhibit a pronounced dependence on detuning, but also that the two locking axes respond differently to a common resonance shift. The latter condition ensures that a transfer operation calibrated at a reference detuning becomes imperfect when the resonance frequency changes.

There is considerable freedom in selecting the two configurations, which need not belong to the same family of pulse sequences. Here, we restrict our attention to the protocol in the main text and its permutations. In particular, we select two configurations related by reversal of the constituent sub-blocks, as illustrated in Fig. S1a. Block A consists of the sequence W4–W1–W2–W3, whereas block B follows the reversed ordering W3–W2–W1–W4, with appropriately chosen carrier-frequency offsets.

This particular choice introduces a useful symmetry. As shown in Fig. S1b, the effective locking fields lie in the *yz* plane and are mirror images of one another with respect to the *xz* plane. Consequently, their magnitudes are identical, while their orientations respond oppositely to changes in detuning. The symmetry also produces identical longitudinal and transverse relaxation times for the two configurations, as confirmed by the calculated curves in Figs. S1c–S1f. These properties considerably simplify the analysis: both configurations exhibit the same intrinsic relaxation dynamics, whereas their distinct angular responses provide the differential sensitivity required for imperfect population transfer.

The choice of nominal detuning is further guided by the relaxation disparity. Ideally, the longitudinal relaxation time should be sufficiently long to accommodate many transfers, while transverse components generated by an imperfect transfer should decay rapidly before the next switching event. We therefore seek a region where $T_{1\delta}$ is large compared with $T_{2\delta}$. We identify such regions via direct Monte Carlo simulations of the stochastic dynamics (Figures S1e and S1f).

These considerations establish the two basic ingredients of the sensing protocol: symmetry-related blocks with identical intrinsic relaxation properties and a differential angular response to detuning. The precise nominal offsets, transfer pulses, and operating point are determined next, taking into account both the relaxation difference and the sensitivity of the accumulated population loss.

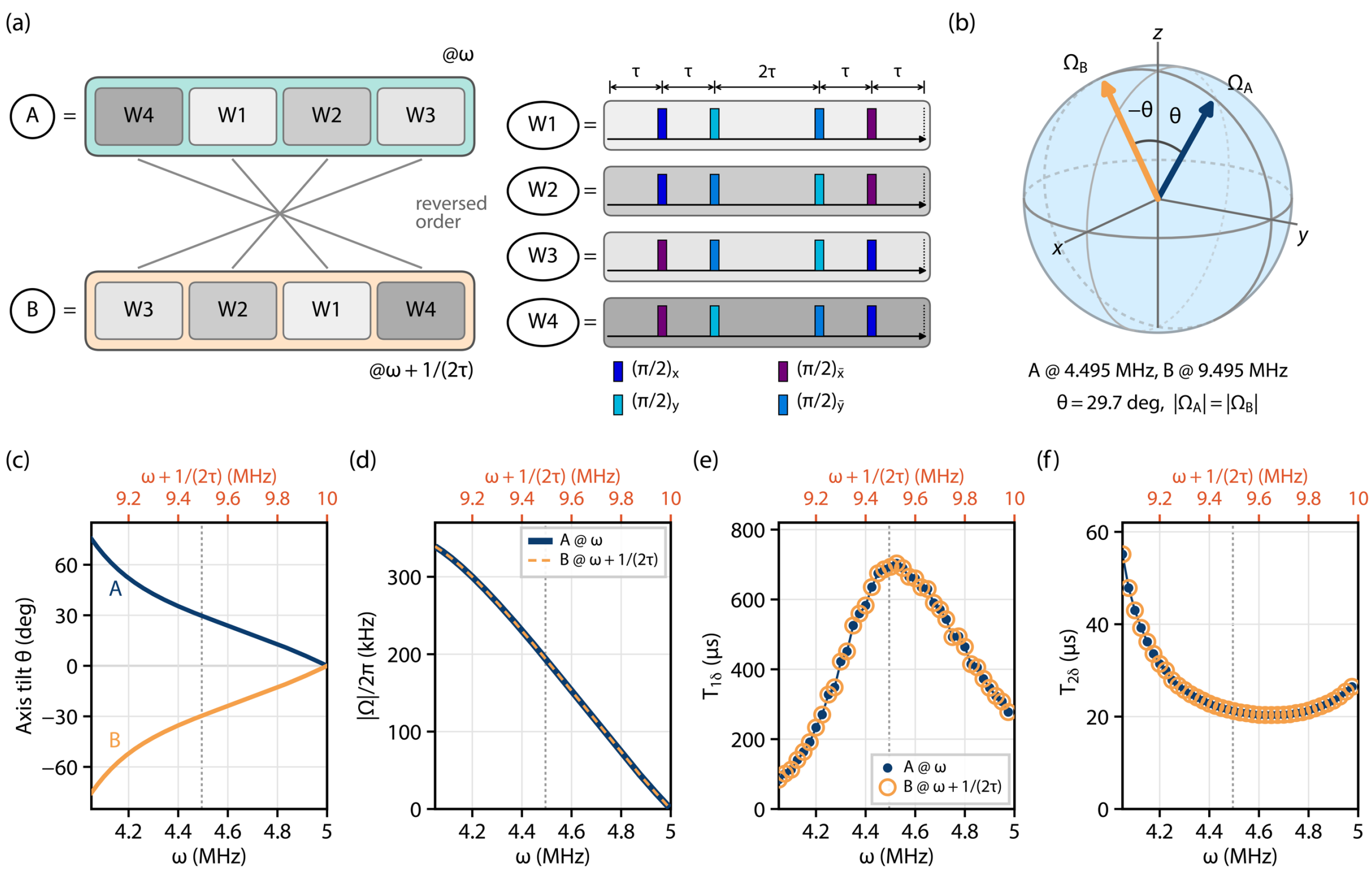

**Fig. S1: Individual properties of the chosen blocks**. A single DSL-4 block as an effective spin lock, and the symmetry that leaves its relaxation unchanged. (a) Block A = W4-W1-W2-W3, with all pulses at detuning ω, and block B, the same sub-blocks in reversed order, at $\omega + \frac{1}{2\tau}$. (b) Effective Floquet fields $\mathbf{\Omega}_\mathrm{A}$ and $\mathbf{\Omega}_\mathrm{B}$ at $\omega =$ 4.495 MHz (dotted lines in (c)-(f)). Both lie in the y–z plane, are mirror images with respect to the *xz* plane, and have equal magnitude. (c) Tilt $\theta$ of the effective axis from *z*, (d) effective rotation frequency $\frac{|\Omega|}{2\pi}$, and (e) $T_{1\delta}$ and $T_{2\delta}$. (f) $T_{2\delta}$ as functions of ω. The top axis gives the detuning $\omega + \frac{1}{2\tau}$ of block B. Symbols in (e) and (f) are Monte Carlo simulations with Lorentzian frequency noise of rms 70.7 kHz and correlation time 1.59 μs; in all calculations, τ = 100 ns.

### II.2 Construction and optimization of the DC sensing protocol

Having selected blocks A and B, we next define the transfer operations and the operating conditions of the DC sensing protocol. An important distinction is made between the "calibration point", at which population transfer between the two locking manifolds is exact, and the "working point", where a controlled transfer mismatch is deliberately introduced to provide a linear response to small resonance shifts.

We first choose the nominal offsets of blocks A and B such that their locking fields, $\mathbf{\Omega}_\mathrm{A}$ and $\mathbf{\Omega}_\mathrm{B}$, lie within the high-anisotropy region identified above. At the calibration point, $\Delta\omega = 0$, transfer pulses are constructed to map the locking axis of A exactly onto that of B, and conversely from B back to A (Fig. S2b, top). The corresponding mismatch angle therefore vanishes, $\epsilon = 0$, and repeated A–B transfers introduce no projective population loss. The residual decay at this point is consequently set by the intrinsic lifetime of the Floquet-locked population.

A common detuning shift $\Delta\omega$ changes the locking axes of both blocks while leaving the calibrated transfer operations unchanged. Because the two axes do not evolve identically with detuning, the transfer is no longer exact and a mismatch angle $\epsilon(\Delta\omega)$ develops. The component projected along the new locking axis is retained, whereas the transverse component relaxes on the shorter $T_{2\delta}$ timescale. Repeated transfers therefore convert the detuning-dependent mismatch into an additional population-decay channel. This behavior is apparent in Fig. S2c, where the decay rate is smallest near the calibration point and increases as the transfer mismatch grows.

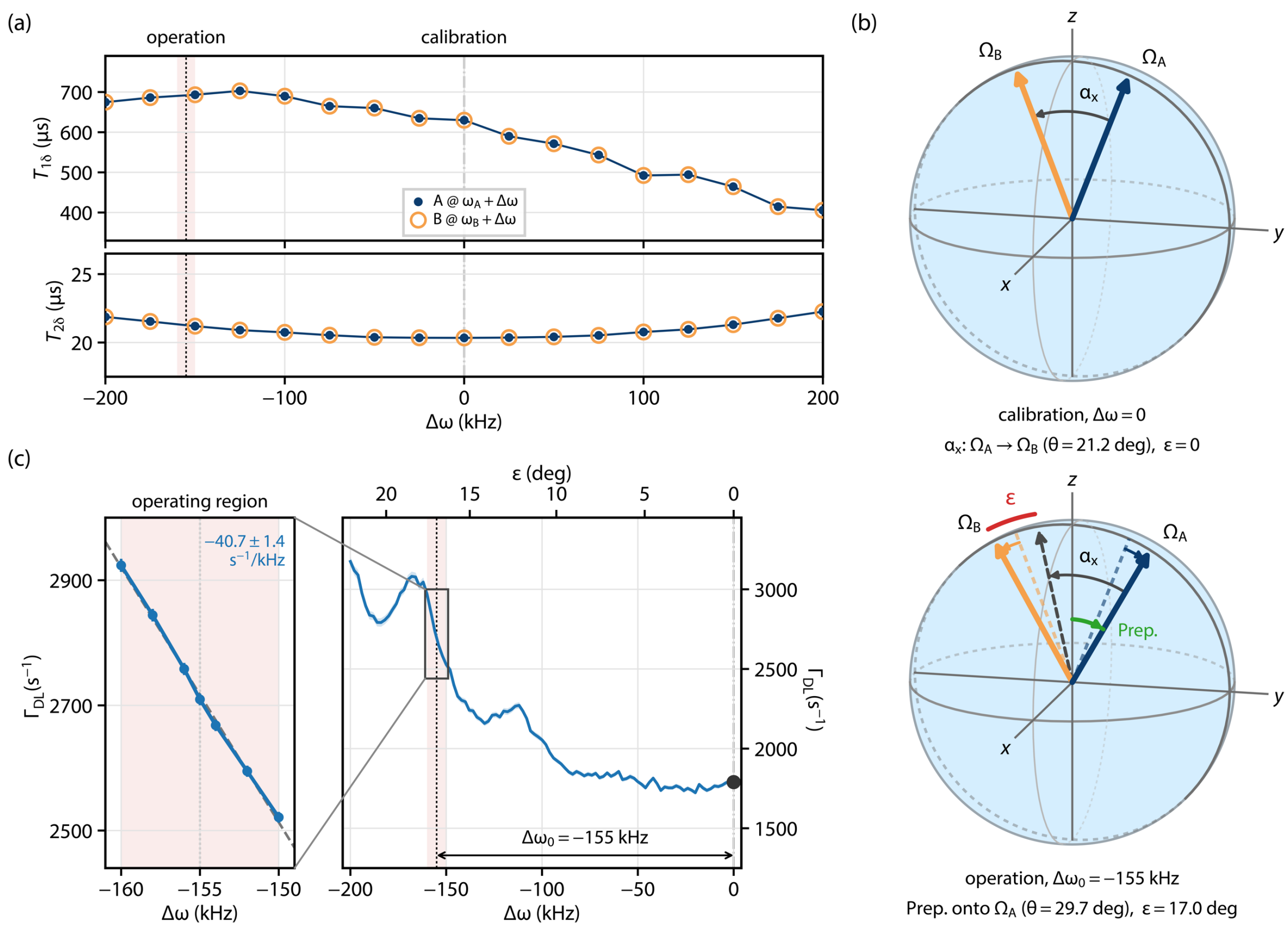

**Fig. S2**: **Construction of the DC protocol**. (a) $T_{1\rho}$ and $T_2$ of block A at $\omega_A - \Delta\omega$ and block B at $\omega_B - \Delta\omega$ versus the common shift $\Delta\omega$, with $\omega_A = 4.650$ MHz and $\omega_B = \omega_A + \frac{1}{2\tau} = 9.650$ MHz. The transfer pulses are calibrated at $\Delta\omega = 0$ (dash-dotted line), and the protocol operates at $\Delta\omega_0 = 155$ kHz (dotted line; shaded, the ±5 kHz operating region), next to the maximum of $T_{1\rho}$. (b) Effective fields at the calibration point (top) and at the operating point (bottom). At $\Delta\omega = 0$, $\alpha_x = 42.35°$ rotates $\mathbf{\Omega}_\mathrm{A}$ exactly onto $\mathbf{\Omega}_\mathrm{B}$. At $\Delta\omega = 155$ kHz both axes have tilted further from $z$, from $\pm 21.2°$ to $\pm 29.7°$(dashed: their positions at $\Delta\omega = 0$). The preparation pulse (Prep., green) is defined at the operating point, so the state starts along the new $\mathbf{\Omega}_\mathrm{A}$. The transfer pulse is not: the same $\alpha_x$ leaves the state at $\epsilon = 17.0°$ from $\mathbf{\Omega}_\mathrm{B}$ (dashed arrow). The return pulse $\beta_x = 2\pi - \alpha_x$ has the same mismatch. (c) Decay rate $\Gamma_\mathrm{DL}$ of the train, with the transfer mismatch $\epsilon$ on the top axis. $\Gamma_\mathrm{DL}$ is lowest near the calibration point (circle), where the transfer is exact, and grows with $\epsilon$. Left: zoom on the operating region, $\Delta\omega = 155 \pm 5$ kHz. There $\Gamma_\mathrm{DL}$ is linear in the detuning.

Operating exactly at the calibration point, however, is not optimal for sensing. Since the transfer is perfect there, the population loss is locally symmetric with respect to detuning and therefore has no first-order response to a small perturbation. We instead introduce a deliberate common bias, shifting both blocks away from their calibration conditions. The working point is chosen near the center of the region of maximum relaxation contrast, where $T_{1\delta}$ remains long while $T_{2\delta}$ is short. This choice simultaneously permits many successive transfers and ensures rapid removal of the transverse component generated after each imperfect transfer. For the parameters used here, the resulting working point corresponds to $\Delta\omega = -155$ kHz, where the transfer already carries a finite mismatch, and the decay rate varies linearly over a reasonably wide interval (Fig. S2c).

The preparation and readout operations are defined at this working, rather than calibration, point (Fig. S2b, bottom). Thus, in the absence of the perturbation to be measured, the initial magnetization is aligned with the appropriate locking axis and read out with maximum contrast, even though the subsequent AB evolution intentionally contains a finite projective loss. A small additional resonance shift then modifies the transfer mismatch and, consequently, the accumulated decay rate. In this way, the imposed bias converts the intrinsically quadratic response around the calibration point into the locally linear sensing response required for DC detection.

Finally, the number $K$ of elementary Floquet cycles within each A or B block must be chosen sufficiently large that transverse magnetization generated by one transfer decays before the next transfer occurs. Otherwise, residual transverse components can be rotated back toward a locking axis at later stages and the simple projective-loss picture no longer applies. At the same time, unnecessarily large $K$ reduces the number of transfers that can be accumulated within the locked-state lifetime. We therefore choose $K = 20$, providing sufficient separation between the $T_{2\delta}$-limited settling and successive transfer events while retaining a large number of AB repetitions within the available $T_{1\delta}$ window.

**II.3 Extension to AC sensing through a symmetry-related protocol**

The DC protocol developed above can be extended to AC sensing by constructing a second protocol with the same relaxation properties but the opposite response to a resonance shift. We denote the original AB protocol by S and its symmetry-related counterpart by T. Alternating between S and T then provides an additional level of temporal modulation: an oscillatory perturbation whose sign reverses synchronously with the S ↔ T alternation produces mismatch-induced loss in the same sense during both portions of the sequence, whereas perturbations at other frequencies are progressively suppressed.

For the pulse sequence considered here, T follows directly from a simple symmetry transformation of S. As illustrated in Fig. S3a, the phase of every RF pulse and the sign of every detuning are reversed. At the level of the DSL-4 sub-blocks, this transformation exchanges W1 ↔ W2 and W3 ↔ W4, such that the sequence defining block A is transformed into B and vice versa. The transfer pulses, which correspond to rotations about *x*, remain unchanged. Thus, the mirrored protocol can be generated from the same basic control elements without requiring an independently optimized set of transfer operations.

The geometric origin of this property is shown in Fig. S3b. All relevant Floquet locking axes lie in the *yz* plane, and the transformation from S to T is equivalent to a $\pi$-rotation about *x*. The locking axes are therefore reversed in sense but preserve the relative geometry relevant for population transfer. Because the transfer rotations are themselves about *x*, they commute with this transformation. A pulse that exactly connects the two locking manifolds of S at the calibration point consequently performs the corresponding exact transfer for T, and the same remains true for the return pulse. Away from calibration, the two protocols acquire equal transfer mismatches while retaining the symmetry-related geometry shown in Fig. S3b. The reversal of the locking-axis sense is immaterial for the spin-locking dynamics, allowing preparation and readout along the same physical axis.

The consequence for sensing is evident in Fig. S3c. The decay-rate responses of S and T are mirror images about the calibration condition. We therefore operate the two protocols at opposite bias points, where their local slopes have equal magnitude and opposite sign. A positive resonance shift increases the mismatch-induced decay rate for one protocol while decreasing it for the other. Conversely, changing the sign of the perturbation exchanges these responses. This antisymmetry provides the ingredient required to rectify an oscillatory perturbation into a cumulative population loss.

Frequency selectivity is then introduced by concatenating the two protocols into higher-order patterns $S^L T^L$, where $L$ specifies the number of consecutive supercycles spent under each sensing condition (Fig. S3d). When the sign reversals of an AC perturbation coincide with the S ↔ T transitions, each half-cycle drives the system toward increased mismatch-induced relaxation: the field changes sign at precisely the same time as the slope of the sensing response. The otherwise alternating perturbation is thereby effectively rectified, allowing its contribution to accumulate throughout the locked-state lifetime. Changing $L$ changes the duration of the S/T modulation and hence shifts the frequency at which this condition is satisfied, as illustrated schematically in the lower panel of Fig. S3d. By contrast, repetition of S alone retains the DC response discussed in the preceding section.

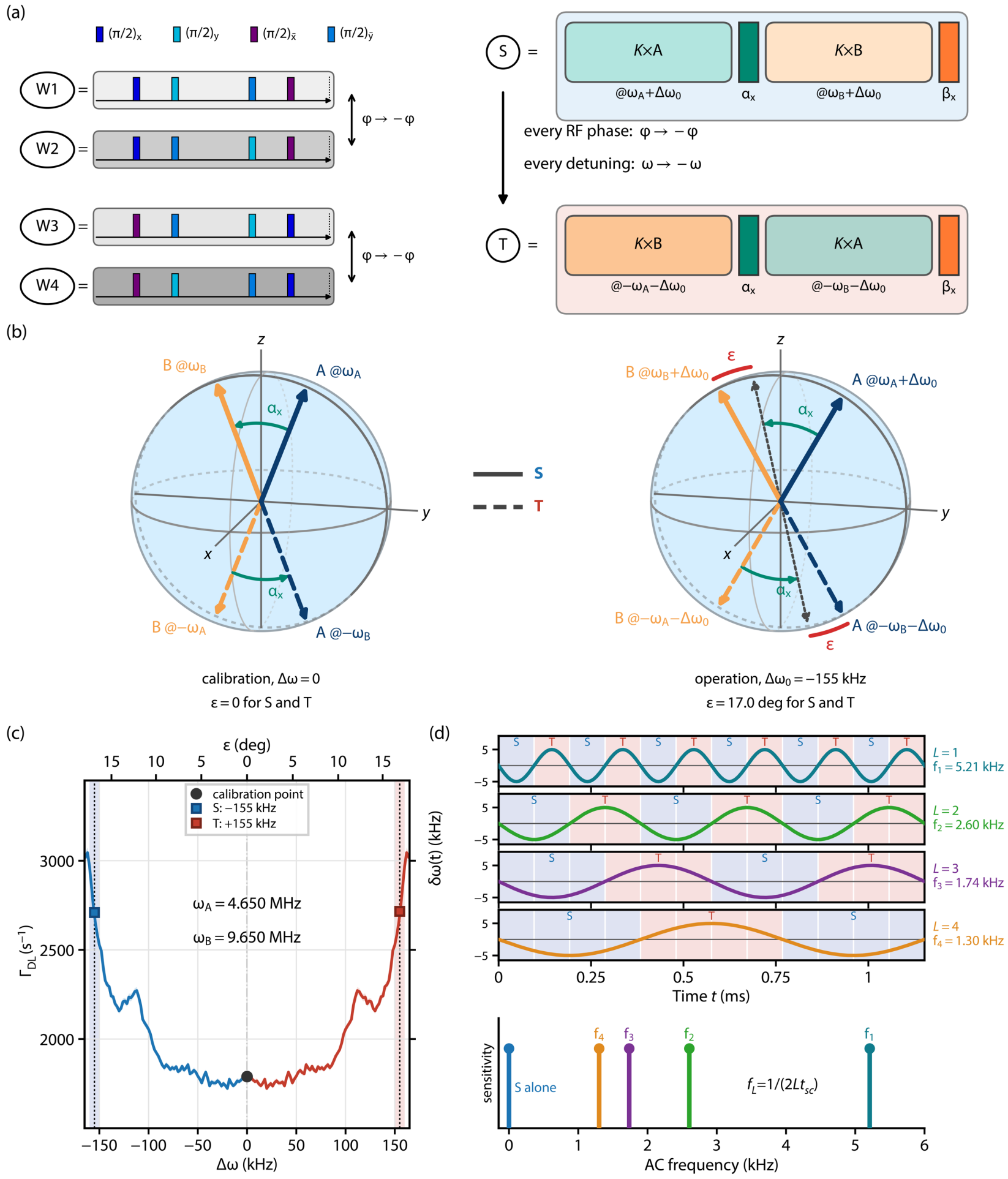

**Fig. S3**: **The mirror protocol T and its use for AC sensing**. (a) Construction of T: every RF phase and every detuning of S changes sign. On the sub-blocks, $\varphi \to -\varphi$ exchanges W1 ↔ W2 and W3 ↔ W4, so block A becomes B, and vice versa. The transfer pulses $\alpha_x$ and $\beta_x$ remain unchanged. (b) Effective axes of the blocks at the calibration point (left) and at the working point (right), colored by block (A, dark blue; B, orange): solid for S and dashed for T. All axes lie in the *yz* plane, so the π-rotation about *x* that maps S onto T reverses each of them. The first block of T, B at $-\omega_A - \Delta\omega$, has the axis of A at $\omega_A + \Delta\omega$with the opposite sense, and the second, A at $-\omega_B - \Delta\omega$, that of B at $\omega_B + \Delta\omega$. The transfer pulses are rotations about *x*, so they commute with that π rotation. Hence $\alpha_x$ (green arcs) carries the axis of the first block onto that of the second in T exactly as it does in S, and $\beta_x = 2\pi - \alpha_x$ closes the supercycle in both.This holds at every $\Delta\omega$: at the working point, the transfer mismatch is the same for S and T, $\epsilon = 17°$. Only the sense of each axis changes, which is immaterial for spin locking, so T is prepared and read out along the same line as S. (c) Decay rate $\Gamma_{DL}$ of S and T versus the common shift $\Delta\omega$ from the calibration point of the transfer pulses (circle), with the transfer mismatch $\epsilon$ on the top axis. S operates at $\Delta\omega = 155$ kHz and T at $-\Delta\omega$ (dotted lines; shaded, the ±5 kHz operating regions). (d) Top: the interleaved patterns ($S^L T^L$ along the 12-supercycle train, supercycle period $t_{sc} = 96$ µs, $L = 1$– 4, each with an AC field of amplitude 5 kHz in phase with the pattern). Bottom: schematic of the resulting frequency selectivity. Each pattern responds at $f_L = \frac{1}{2Lt_{sc}}$, whereas S alone responds to a static shift. The calculated response is shown in Fig. 3d in the main text.

This construction separates two functions of the control protocol that are useful to keep conceptually distinct. The internal A/B transfers determine the parameter-dependent relaxation and its cumulative amplification, whereas the slower alternation between S and T determines the frequency selectivity. The latter can therefore be adjusted without changing the underlying Floquet locking mechanism, providing the basis for the AC sensing protocol presented in the main text.

## III. Sensitivity and long-term precision

The analysis in Section I distinguishes two implementations of relaxation-assisted sensing: estimation from the population at a selected interrogation time and estimation from the relaxation rate itself. We now examine the latter under a detection model representative of a single NV center. Our goal is not to revisit the intrinsic sensitivity derived above, but to determine how the two sensing modalities behave over long acquisition times when the optical contrast is not perfectly stationary. Indeed, long-duration measurements with solid-state spin sensors are affected by slow variations in fluorescence intensity, optical collection efficiency, initialization and readout fidelity, and other experimental parameters. These fluctuations are particularly important for NV-center measurements because the spin state is inferred from a fluorescence contrast. Consequently, a change in detected contrast (stemming, e.g., from photo-induced changes in the NV charge state cyclicity, a common scenario near the diamond surface) can become indistinguishable from a change in spin population when the sensing observable is encoded in a single intensity measurement. Here we introduce a simple detection model incorporating such fluctuations and use it to compare the long-term precision of conventional Ramsey and relaxation-based Floquet sensing.

### III.1 Model for NV-spin detection and contrast drift

We model the optical detection of a single NV center by combining the spin responses calculated for the Ramsey and Floquet protocols with a statistical model of NV fluorescence detection. The spin responses used here are obtained using the same Lorentzian frequency-noise model employed throughout our numerical calculations, with a spectral density centered at zero frequency, a FWHM of 200 kHz, and an rms amplitude of 70.7 kHz. In the present section these spin trajectories are taken as fixed inputs: we do not resimulate the spin dynamics for each realization of the measurement. Instead, the simulations described below generate the photon-counting statistics and slow variations in optical contrast expected during repeated measurements of those underlying responses.

For a single optical readout, we assume an average photon yield of 0.03 photons for an $NV^-$ center in $m_S = 0$, an initial spin contrast 0.30, and a probability 0.25 that the center is in the neutral $NV^0$ charge state at the time of readout. The $NV^0$ fluorescence is assigned a relative brightness 0.30. Together with the spin population provided by the calculated response, these parameters determine the mean photon number and its variance for each readout. At the level of an individual measurement the photon statistics are Poissonian. Because the simulations considered here involve very large numbers of repetitions, however, the accumulated photon counts are drawn directly from the corresponding Gaussian distribution.

Both sensing protocols are implemented with two-phase readout to reject fluorescence contributions unrelated to the spin state. For Ramsey sensing, measurements with opposing readout phases are subtracted to obtain the spin-dependent signal. The Floquet protocol uses an analogous procedure: consecutive measurements project along $+\mathbf{\Omega}_A$ and $-\mathbf{\Omega}_A$, and their difference provides each point of the relaxation trajectory. The important quantity for the present analysis is the contrast, since it provides the conversion between the underlying spin population and the experimentally measured optical signal. We allow this conversion factor to drift slowly in time. Specifically, the simulations are organized into measurement batches of duration $T_b = 46.552$ ms, chosen such that each batch contains either 9800 Ramsey repetitions or 44 complete Floquet relaxation traces. Each Floquet trace comprises the 11 sampling points. We assume that the optical contrast is effectively constant over one batch but may vary between successive batches. This coarse graining is convenient computationally: rather than drawing the photon number for every individual shot, we accumulate the expected Poisson statistics over the repetitions contributing to each measured point. For the large number of repetitions considered here, the resulting distribution is well approximated by a Gaussian with the corresponding mean and variance. Thus, a batch represents the shortest interval over which we explicitly update the slowly varying experimental contrast.

We model this contrast as $\mathcal{C}(t) = \mathcal{C}_0\big(1+\zeta(t)\big)$, where $\mathcal{C}_0$ is the nominal spin contrast and $\zeta(t)$ is a zero-mean, slowly varying fractional fluctuation. To represent the long-timescale instrumental drift, $\zeta(t)$ is generated with a $1/f$ power spectral density. Its amplitude is specified through the Allan deviation $\sigma_{\mathcal{C}}$, defined from the fluctuations of the average contrast in consecutive windows of duration $T_b$. For $1/f$ noise, $\sigma_{\mathcal{C}}$ is independent of the averaging time, while the rms excursion relative to the initial calibration grows with the elapsed measurement time $t$ as $rms\big(\zeta(t)\big) = \sigma_{\mathcal{C}}\sqrt{\frac{\ln(t/T_b)}{\ln 2}}$. We use $\sigma_{\mathcal{C}} = 0.05$ throughout the simulations of Fig. S4. Each batch $j$ is therefore assigned a single multiplicative contrast factor $\big(1+\zeta_j\big)$, common to all Ramsey shots or Floquet readouts acquired within that batch, while $\zeta_j$ evolves in a temporally correlated manner from one batch to the next. The first batch defines the contrast calibration, so that the subsequent drift is referenced to its initial value.

This common-mode variation within each batch affects the two sensing strategies differently. For Ramsey sensing, the resonance shift is inferred directly from a measured population contrast. A slow multiplicative change in contrast therefore produces an apparent change in resonance frequency and cannot, without an independent normalization measurement, be distinguished from the signal of interest. Figure S4b, top, illustrates this effect for a true frequency shift $\delta\omega = 5$ kHz: the estimated frequency acquires a slowly varying contribution correlated with the contrast drift.

For Floquet sensing, by contrast, each measurement batch contains a set of measurements sampling the relaxation trajectory, $S_{DL} = \mathcal{V}e^{-\Gamma_{DL}(\omega)t_k} + \mathcal{W}$, where $\mathcal{V}$ and $\mathcal{W}$ are treated as nuisance parameters and $\Gamma$ is the quantity from which the resonance shift is inferred. Because the contrast remains approximately constant over this interval, a slow fluctuation primarily rescales $\mathcal{V}$ and $\mathcal{W}$ while leaving the fitted $\Gamma_{\mathrm{DL}}$ unchanged. The relaxation rate is subsequently converted into a resonance shift using the locally calibrated relation.

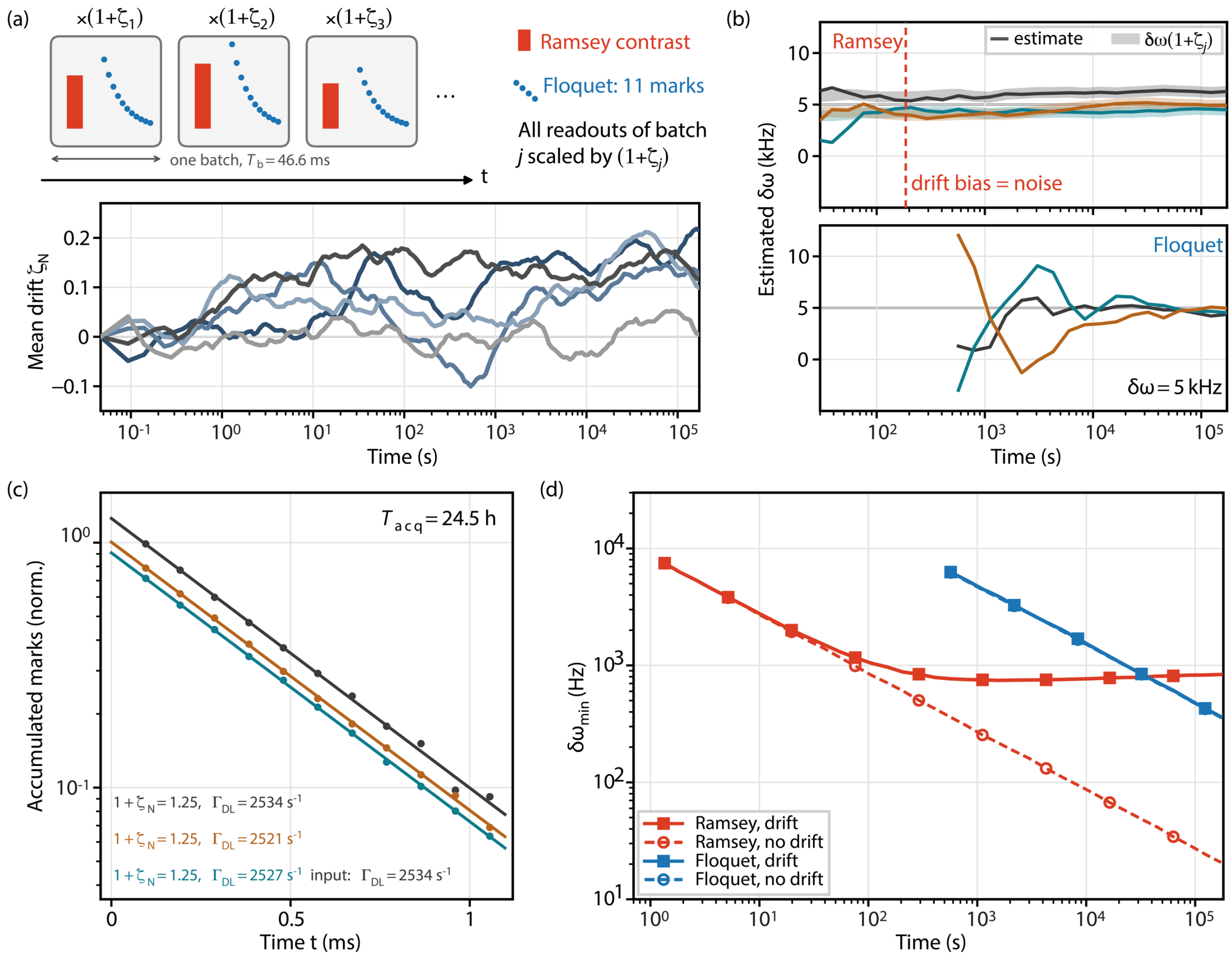

**Fig S4**: **Effect of slow contrast drift on long-term sensing precision.** (a) Model for temporally correlated contrast fluctuations. Each measurement batch experiences a multiplicative factor $(1+\zeta)$, common to all measurements within the batch but slowly varying between batches. Ramsey sensing uses the population contrast, whereas Floquet sensing extracts the relaxation rate from the temporal evolution within each batch. Representative realizations of the fractional drift $\zeta$ are shown below, where we defined $\zeta_N \equiv \frac{1}{N}\sum_{j=1}^{N}\zeta_j$ and $N$ is the number of accumulated measurement batches. (b) Resonance-frequency estimates obtained from Ramsey (top) and Floquet relaxation-rate sensing (bottom) for a fixed shift $\delta\omega = 5$ kHz. Slow contrast fluctuations produce a systematic contribution to the Ramsey estimate, while their effect is largely rejected by the rate-based measurement. (c) Simulated Floquet relaxation traces for different multiplicative contrast factors. Although the signal amplitude changes, the fitted relaxation rate remains essentially unchanged. (d) Minimum resolvable frequency shift versus total averaging time for Ramsey and Floquet sensing, with and without contrast drift. In the absence of drift, Ramsey retains the higher short-term sensitivity; in the presence of slow fluctuations its resolution reaches a floor, whereas Floquet sensing continues to improve with averaging and eventually provides superior long-term precision.

Figure S4c isolates this effect by applying different multiplicative contrast factors to otherwise identical Floquet relaxation traces. The amplitudes of the measured trajectories change appreciably, whereas the extracted rates remain essentially unchanged. Thus, provided the contrast fluctuations are quasistatic over a single relaxation trace, the temporal shape of the signal carries information that is largely independent of its absolute magnitude. This is the practical realization of the rate-based sensing principle introduced in Section I.3.

### III.2 Long-term precision under slow instrumental fluctuations

To compare the two sensing strategies over experimentally relevant averaging times, we generate repeated measurement batches over a total acquisition time $T_{\mathrm{acq}}$ and estimate the resonance shift from the accumulated data. The true shift is fixed at $\delta\omega = 5$ kHz (or the equivalent of ~180 nT for an electron spin). For Ramsey sensing, the accumulated differential fluorescence signal is converted into a frequency shift using the calibrated Ramsey response. For Floquet sensing, the accumulated counts at the 11 sampling times are used to reconstruct the relaxation trajectory, which is fitted to extract $\Gamma_{DL}$; the corresponding frequency shift is then obtained from the calibrated dependence $\Gamma_{DL}(\omega)$. For each acquisition time, this procedure is repeated over $10^4$ independent realizations of the complete measurement, each with an independently generated contrast-drift history and photon-counting noise. We define the frequency resolution $\delta\omega_{min}$ as the standard deviation of the estimation error $\delta\omega_{est} - \delta\omega_{true}$ over this ensemble. This definition therefore includes both statistical uncertainty and the realization-dependent bias introduced by contrast drift. In particular, $\delta\omega_{min}$ should not be confused with an Allan deviation: the latter is evaluated from fluctuations within a single temporal record and can remain insensitive to a slowly varying or persistent bias, whereas the present metric quantifies the accuracy with which the true resonance shift is recovered across independent realizations of the complete experiment.

The resulting comparison is shown in Fig. S4d. In the absence of slow contrast drift, Ramsey sensing provides the superior short-term sensitivity and continues to improve approximately with increasing averaging time. This is consistent with the statistical penalty associated with estimating a relaxation rate from several temporal samples rather than extracting the parameter from a single optimally chosen population measurement. When contrast drift is included, however, the Ramsey resolution eventually reaches a floor: additional averaging no longer improves the estimate because the slowly varying systematic contribution is not removed by repeated measurements. In contrast, the Floquet rate estimate remains largely insensitive to the same fluctuations and continues to improve as additional relaxation trajectories are accumulated. A crossover therefore occurs beyond which the relaxation-based protocol provides the better frequency resolution, despite its poorer ideal short-term sensitivity.